\documentclass[twocolumn]{aa}  
\usepackage{makecell}
\usepackage{graphicx}
\usepackage{multirow}
\usepackage{txfonts}
\newcommand{\quotes}[1]{``#1''}
\begin{document} 

\title{The ionizing properties of two bright Ly$\alpha$ emitters in the BDF reionized bubble at z=7}
   \author{M. Castellano\inst{1}, L. Pentericci\inst{1}, G. Cupani\inst{2}, E. Curtis-Lake\inst{3}, E. Vanzella\inst{4},  R. Amor\'in\inst{5,6}, D. Belfiori\inst{1}, A. Calabr\`o\inst{1}, S. Carniani\inst{7}, S. Charlot\inst{8}, J. Chevallard\inst{9}, P. Dayal\inst{10}, M. Dickinson\inst{11}, A. Ferrara\inst{7}, A. Fontana\inst{1}, E. Giallongo\inst{1}, A. Hutter\inst{10}, E. Merlin\inst{1}, D. Paris\inst{1}, P. Santini\inst{1}}

   \institute{
              $^{1}$INAF -- OAR, via Frascati 33, 00078 Monte Porzio Catone (Roma), Italy\\
              $^{2}$INAF -- Osservatorio Astronomico di Trieste, Via G. B. Tiepolo 11, I-34143 Trieste, Italy\\
              $^{3}$Centre for Astrophysics Research, Department of Physics, Astronomy and Mathematics, University of Hertfordshire, Hatfield, AL10 9AB, UK\\
              $^{4}$INAF -- OAS, Osservatorio di Astrofisica e Scienza dello Spazio di Bologna, via Gobetti 93/3, I-40129 Bologna, Italy\\
              $^{5}$Departamento de Astronomía, Universidad de La Serena, Av. Juan Cisternas 1200 Norte, La Serena, Chile\\
              $^{6}$Instituto de Investigaci\'on Multidisciplinar en Ciencia y Tecnolog\'ia, Universidad de La Serena, Ra\'ul Bitr\'an 1305, La Serena, Chile\\
              $^{7}$Scuola Normale Superiore, Piazza dei Cavalieri 7, I-56126 Pisa, Italy\\
              $^{8}$Sorbonne Universit\'e, CNRS, UMR7095, Institut d'Astrophysique de Paris, F-75014, Paris, France\\
              $^{9}$Astrophysics, The Denys Wilkinson Building, University of Oxford, Keble Road, Oxford, OX1 3RH\\
              $^{10}$Kapteyn Astronomical Institute, University of Groningen, P.O. Box 800, 9700 AV Groningen, The Netherlands\\
              $^{11}$NSF’s NOIRLab, Tucson, AZ 85719, USA\\
             }

   \date{...}

% \abstract{}{}{}{}{} 
% 5 {} token are mandatory
 
  \abstract
   {}
   {We investigate the ionizing properties of the pair of bright Ly$\alpha$ emitting galaxies BDF521 and BDF2195 at z=7.012 in order to constrain their contribution to the formation of the BDF "reionized bubble" \citep[][]{Castellano2016,Castellano2018} in which they have been shown to reside.
   }
   {We obtain constraints on UV emission lines (CIV$\lambda 1548$ doublet, HeII$\lambda 1640$, OIII]$\lambda 1660$ doublet, and CIII]$\lambda 1909$ doublet) from deep VLT-XSHOOTER observations and compare them to those available for other high-redshift objects, and to models with mixed stellar and AGN emission. We use this spectroscopic information together with the photometry available in the field to constrain the physical properties of the two objects using the spectro-photometric fitting code \textsc{BEAGLE}.}
   {We do not detect any significant emission at the expected position of UV lines, with 3$\sigma$ upper limits of EW$\lesssim$2-7\AA~rest-frame. We find that the two objects have lower CIII] emission than expected on the basis of the correlation between the Ly$\alpha$ and CIII] EWs. The EW limits on CIV and HeII emission exclude pure AGN templates at $\sim2-3\sigma$ significance, and only models with a $\lesssim$40\% AGN contribution are compatible with the observations. The two objects are found to be relatively young ($\sim$20-30 Myrs) and metal-poor ($\lesssim 0.3 Z_{\odot}$) with stellar masses of a few $10^9M_{\odot}$. Their production rate of hydrogen ionizing photons per intrinsic UV luminosity is log($\xi_{ion}^*$/Hz erg$^{-1}$)=25.02-25.26, consistent with values typically found in high-redshift galaxies, but more than twice lower than values measured in $z>$7 galaxies with strong CIII] and/or optical line emission ($\simeq$25.6-25.7).}
   {The two BDF emitters have no evidence of higher than average ionizing capabilities and are not capable of reionizing their surroundings by their own means under realistic assumptions on the escape fraction of ionizing photons. Therefore, a dominant contribution to the formation of the reionized bubble must have been provided by companion fainter galaxies. These objects will need JWST capabilities for spectroscopic confirmation.}
\authorrunning{M. Castellano et al.}   
\titlerunning{The ionizing properties of two bright Ly$\alpha$ emitters in the BDF bubble}   

\maketitle
%
%________________________________________________________________

\section{Introduction}

The epoch of reionization (EoR) marked a major phase transition of the Universe, during which the intergalactic medium (IGM) became transparent to UV photons. Determining the \textit{physical processes} involved in the reionization process, its \textit{timeline} and \textit{topology} represents the latest frontier in observational cosmology \citep[][]{Dayal2018}. 

The first indication of an increased neutral hydrogen fraction  ($\chi_{HI}$) in the IGM at $z\gtrsim$6 was obtained by observations of the Gunn-Peterson effect in distant quasars \citep[][]{Fan2002}. However, a  substantial step in our knowledge of the reionization \textit{timeline} has been made possible only by constraints on the redshift evolution of the fraction of Lyman-break galaxies (LBGs) that have an appreciable Ly$\alpha$ emission line \citep[e.g.,][]{Malhotra2006,Stark2010,Fontana2010}. In fact, the neutral hydrogen in the IGM scatters Ly$\alpha$ photons out of the line-of-sight, such that a lower line visibility indicates a higher IGM neutral fraction if the properties of the inter-stellar medium remain unchanged \citep[e.g.,][]{Dayal2011,Dijkstra2019}. As of today, a decrease in the Ly$\alpha$ fraction at $z\gtrsim$6 has been confirmed by many independent analyses and interpreted as indication of an increasing neutral hydrogen fraction in the IGM \citep[e.g.,][]{Pentericci2011,Schenker2012,Ono2012,Mason2018, Pentericci2018,Mason2019}. Together with measurements of the CMB Thomson optical depth \citep[][]{Planck2020}, of the clustering of Ly$\alpha$-emitters \citep[LAEs,][]{Hutter2015,Sobacchi2015} and with measurements of ionized regions around the most distant QSOs \citep[e.g.,][]{Banados2018}, the available observations point to a scenario in which the universe was still substantially neutral at $z\sim$10 ($\chi_{HI}\gtrsim$90\%) and was rapidly reionized in $\sim$500Myrs, ending the EoR at $z\sim$5.5-6 \citep[e.g.,][]{Mitra2015,Greig2017,Mitra2018,Hutter2021}. Star-forming galaxies are currently considered as the most likely responsible of reionization \citep[e.g.,][]{Robertson2015,Bouwens2015b,Dayal2020,Endsley2021a,Romanello2021}, although our poor knowledge of galaxy physical properties does not allow us to rule out that active galactic nuclei (AGN) also contributed \citep[][]{Finkelstein2019,Giallongo2015}. 

The rate of ionizing photons escaping into the IGM from a given population is $\dot{N}=\rho_{UV}\xi_{ion}^*f_{esc}$, where $\rho_{UV}$ is the total, dust-corrected UV luminosity density at 1500\AA,  $\xi_{ion}^*$ is the ionizing photon production efficiency per unit UV luminosity, and $f_{esc}$ is the fraction of ionizing photons leaked into the surrounding environment. Currently, the UV luminosity density from star-forming galaxies is well constrained by measurements of their UV luminosity function (LF) at $z>$6 down to $L<L^*$ \citep[e.g.,][]{Bouwens2015,Livermore2017,Ishigaki2018,Yue2018,Oesch2018}. On the other hand, the escape fraction of ionizing photons can only be directly constrained at $z\lesssim$3-4 and on average is quite modest in LBGs with  values lower than 5-10\% \citep[][]{Boutsia2011,Marchi2017,Grazian2017,Steidel2018,Pahl2021}. Only in few rare objects it reaches values of $f_{esc}\gtrsim$50\% \citep[][]{deBarros2016,Vanzella2016,Naidu2017,Vanzella2018,Izotov2018a,Izotov2018b,Flury2022}. As of today, no real constraints exist on a possible increase of $f_{esc}$ with redshift and we can only make assumptions on its value in the EoR.

The ionizing efficiency $\xi_{ion}^*$ can be directly obtained by measuring, at the same time, the non-ionizing UV continuum and the Balmer emission lines, the latter yielding the emission rate of ionizing photons \citep[e.g.,][]{Shivaei2018}, or from the equivalent width of the [OIII]$\lambda 4959,5007$ doublet \citep[][]{Chevallard2018,Tang2019}. At higher redshifts where optical emission lines are not directly observable through spectroscopy, constraints on $\xi_{ion}^*$ were derived by the analysis of the rest-frame UV colors \citep[e.g.,][]{Duncan2015} or from the contamination from strong emission lines to the mid-IR colors, finding a typical log($\xi_{ion}^*$/Hz erg$^{-1}$)$\simeq$25.3 at $z\sim$4-5 \citep[][]{Bouwens2016a,Lam2019}, with indication of higher ionizing efficiencies in LAEs \citep[][]{Harikane2018,Sobral2019} and strong [OIII]$\lambda 4959,5007$ emitters \citep[][]{Tang2019}.

For galaxies in the EoR, constraints on the ionizing capabilities can be derived using UV emission lines. These however are extremely faint and have been detected only in a handful of objects \citep[e.g.,][]{Stark2015a,Stark2015b,Laporte2017,Mainali2018, Hutchison2019,Topping2021}.

Such features might be associated to AGN or metal poor stellar populations \citep[][]{Gutkin2016,Feltre2016,Nakajima2018a} as also shown at lower redshifts \citep[e.g.,][]{Amorin2017,Calabro2017,Nakajima2018b}.
The spectro-photometric analysis of a CIV emitter at z=7.045 allowed \citet{Stark2015b} to estimate a ionizing efficiency log($\xi_{ion}^*$/Hz erg$^{-1}$)$\simeq$25.68 much higher than the typical value measured at lower redshifts, and comparable to values measured in rare, extreme emission line galaxies at $z\sim$3-4 \citep[][]{Nakajima2016}. Similarly, spectroscopically confirmed $z>$7 galaxies with mid-IR colors suggestive of intense H$\beta$+[OIII] emission show a log($\xi_{ion}^*$/Hz erg$^{-1}$)$\gtrsim$25.5 \citep[][]{Stark2017,Endsley2021b}. These results point to an increased fraction of galaxies with high ionizing efficiency in the EoR.

A promising way to shed light on the sources of reionization is the investigation of regions where this process is more advanced and observations point to the presence of "reionized bubbles" \citep[][]{Castellano2016,Higuchi2019,Tilvi2020,Jung2021,Leonova2021,Endsley2022}. The analysis of the Ly$\alpha$ fraction in independent lines-of-sight suggests that reionization was a spatially inhomogeneous process \citep[][]{Treu2012,Pentericci2014}. A detailed spectroscopic investigation of the Bremer Deep Field (BDF) \citep[][]{Lehnert2003,Castellano2010b}, led to the first discovery of a likely reionized "bubble" at $z\sim$7 \citep[][hereafter C16]{Castellano2016}. The BDF hosts three bright ($L\sim L^*$) Ly$\alpha$ emitting galaxies with EW$>$50\AA~\citep[][hereafter V11 and C18]{Vanzella2011,Castellano2018}, and the density of $z\simeq$7 faint LBGs in this area  is $\gtrsim$3-4 times higher than the average (C16). Two of the Ly$\alpha$ emitting galaxies (BDF521 and BDF2195) have exactly the same  redshift and are at a projected physical separation of only 91.3kpc, the third one, BDF3299, being at 1.9 proper Mpc projected distance. 

The high number density of LBGs in the BDF region is consistent with the clustered faint galaxies being key contributors to the local reionization history \citep[C16,][]{Espinosa2021}, as expected in ``inside-out'' reionization scenarios \citep[][]{Choudhury2009,Dayal2009,Trebitsch2021}. However, significant uncertainties remain due to the lack of Ly$\alpha$ detections in the  faint companion galaxies, despite the significant observational efforts (C18). Lacking a spectroscopic confirmation of the faint LBGs in the field, it is fundamental to fully constrain the physical properties of the bright emitters and assess whether they can create the reionized region possibly thanks to hard-ionizing stellar populations or AGN.

In this paper we analyse deep VLT-XSHOOTER observations of the BDF521 and BDF2195 pair to put constraints on the ionizing budget of the two emitters from UV rest-frame metal emission lines, and to ascertain whether these $L\simeq L^*$ galaxies play a major role in the creation of the BDF bubble.

The paper is organised as follows: in Sect.~\ref{sec:obs} we present the observations and data reduction, in Sect.~\ref{sec:lines} we discuss the constraints on emission lines and compare the BDF galaxies to other high redshift sources. We place constraints on the contribution from AGN in Sect.~\ref{sec:AGN}, and quantify their physical properties and their contribution to the creation of the BDF bubble in Sect.~\ref{sec:ion}. The results are summarised in Sect.~\ref{sec:summary}.

Throughout the paper we adopt AB magnitudes \citep{Oke1983}, a solar metallicity $Z_{\odot}$=0.02, and a $\Lambda$-CDM concordance model ($H_0$ = 70 km s$^{-1}$ Mpc$^{-1}$, $\Omega_M=0.3$, and $\Omega_{\Lambda}=0.7$).

\section{The XSHOOTER observations} \label{sec:obs}

The BDF521 and BDF2195 pair was observed by XSHOOTER in nodding mode with 11$\times$0.9 arcsec slits and 900 and 865 seconds per single exposure in the NIR and VIS arms, respectively. Observations were acquired between 26 September 2019 and 8 June 2021. The total observing time was of 15 and 12 hours for BDF521 and BDF2195, respectively, corresponding to on-target exposure times of 12.0 (11.5) and 9.5 (9.1) hours in NIR (VIS). 

Science frames were reduced with the official XSHOOTER pipeline \citep{Modigliani2010}, v.~3.3.5, using the associated raw calibrations from the ESO archive. Sky subtraction was performed using the XSHOOTER nodding strategy, combining frames acquired at two different positions in the sky for each OB execution. A fixed boxcar window was used to extract the target on the rectified 2D spectra, to ensure that the target was correctly localized along the slit (\texttt{extract\-method=LOCALIZATION} and \texttt{localize\-method=MANUAL} in pipeline recipe \texttt{xsh\_scired\_slit\_nod}). The position of the window with respect to the slit center and its half height (in arcsec) were fine-tuned for the two targets independently, based on the position of the Ly$\alpha$ line which is well detected in the 2D VIS median stacked spectra (and visible in each single OB); we used \texttt{localize\--slit\--position=-1.0}, \texttt{localize\--slit\--hheight=0.5} for BDF521 and \texttt{localize\--slit\--position=-3.2}, \texttt{localize\--slit\--hheight=0.6} for BDF2195 in pipeline recipe \texttt{xsh\_scired\_slit\_nod}.

The 1D extracted spectra, one for each XSHOOTER arm and OB execution, were corrected to the barycentric reference frame in vacuum and combined with the data analysis package \textsc{Astrocook} \citep{Cupani2020}. The combined spectra were rebinned to 18 km s$^{-1}$ in the NIR arm and to 11 km s$^{-1}$ in the VIS arm, roughly corresponding to 3 pixels per resolution element, given the nominal resolution FWHMs of the instrument in the three arms for the adopted slit. Undersampled spectra with larger spectral bins (54 km s$^{-1}$ in the NIR arm, 33 km s$^{-1}$ in the VIS arm) were also produced to facilitate the identification of spectral features by visual inspection. 

While the slits are not designed to include the total flux from the observed objects, following \citet{Lemaux2009} we expect slit losses to be small ($\sim$10-15\%) on the basis of the slit dimension, the compact size of the sources ($\sim$0.15'') and the median seeing ($\sim$0.8'') of the XSHOOTER observations. In order to quantitatively assess the significance of potential slit losses we smoothed the HST F125W image (C16) of the objects to the seeing of the spectroscopic observations and measured the fraction of the total flux observed within the slit size. We find corrections of $\sim$10\% that are, however, uncertain due to effects of light contamination from other objects and local background subtraction of the images. Therefore, we decided not to apply correction factors in the following.

\section{Constraints on UV rest-frame line emission} \label{sec:lines}

At the Ly$\alpha$ redshift estimated from FORS2 (z=7.008, C18), the VIS arm of the instrument covers the Ly$\alpha$ line, while the NIR arm covers the range of the CIV doublet (1548\AA -1550\AA), HeII$\lambda 1640$, the OIII] doublet (1661\AA-1666\AA), and the CIII] doublet (1907\AA-1909\AA). We first visually inspected the spectra to localize Ly$\alpha$ emission and obtain a precise positioning of the object within the slit. 
\subsection{Ly$\alpha$ emission} \label{sec:Lya}

\begin{table}[ht]
\caption{3-$\sigma$ upper limits on the observed fluxes and the rest-frame EW of the UV emission lines of BDF521 and BDF2195}
\centering
\begin{tabular}{| c || c | c || c | c |}
\hline
 \multirow{2}*{Line} & \multicolumn{2}{c| |}{ BDF521}&\multicolumn{2}{c|}{ BDF2195}\\
\cline{2-5}
 & \makecell{Flux \\ $10^{-18}$\\$erg/s/cm^2$} & \makecell{EW \\ \AA} &  \makecell{Flux \\ $10^{-18}$\\$erg/s/cm^2$} & \makecell{EW \\ \AA} \\ \hline \hline
CIV1548.20 & $<$1.40 & $<$7.18 & $<$1.93 &$<$5.77 \\
CIV1550.78 & $<$1.42 &  $<$7.30 & $<$1.79 &  $<$5.58\\
HeII1640.42 & $<$1.47 &  $<$5.86 & $<$1.81 &  $<$4.34\\
OIII1660.81 &  	$<$1.58 &  $<$6.49 & $<$1.80 & $<$3.98\\
OIII1666.15 &  $<$1.58 &  $<$5.96 & $<$2.02  &  $<$4.64\\
CIII1906.68  & $<$1.35 & $<$1.96 & $<$1.81  & $<$1.85\\
CIII1908.73 &  $<$1.45 & $<$2.40& $<$1.87  & $<$2.16\\
\hline
\end{tabular} \label{linestable}
\end{table} 
We detect Ly$\alpha$ line emission for both objects: the flux-weighted mean wavelength of the line corresponds to z=7.0121 and z=7.0124 for BDF521 and BDF2195, respectively. These redshifts are slightly higher than previously estimated from the peak of the Ly$\alpha$ emission in FORS2 spectra (z=7.008) but do confirm that the objects are approximately at the same redshift and, given the close separation on the sky (91.3kpc, C18), likely physically connected. We will use these new estimates in the present analysis. The total Ly$\alpha$ fluxes are $12.8\pm1.0$ and $21.8\pm1.1$ $10^{-18}$ erg/s/cm$^2$ for BDF521 and BDF2195, respectively. The widths in velocity space and the equivalent widths of the lines are consistent with previous analysis. We find EW(Ly$\alpha$)=60$\pm$5\AA, FWHM(Ly$\alpha$)=263$\pm$5 km/s for BDF521, and EW(Ly$\alpha$)=62$\pm$3\AA, FWHM(Ly$\alpha$)=250$\pm$3 km/s for BDF2195. The continuum flux is estimated on the basis of the available photometry (C16). The FWHM is measured with a Gaussian fit and corrected for the effect of instrumental resolution. Uncertainties are computed following \citet{Lenz1992}.

The VIS arm covers also the expected range of the NV$\lambda$ 1240 doublet but in a noisy, low efficiency, region of the spectrum that does not allow us to improve upon the limits previously obtained with FORS2 (C18). 
\begin{figure*}
\centering
\includegraphics[trim={1cm 2cm 1cm 0.5cm},clip,width=\textwidth,height=\textheight,keepaspectratio]{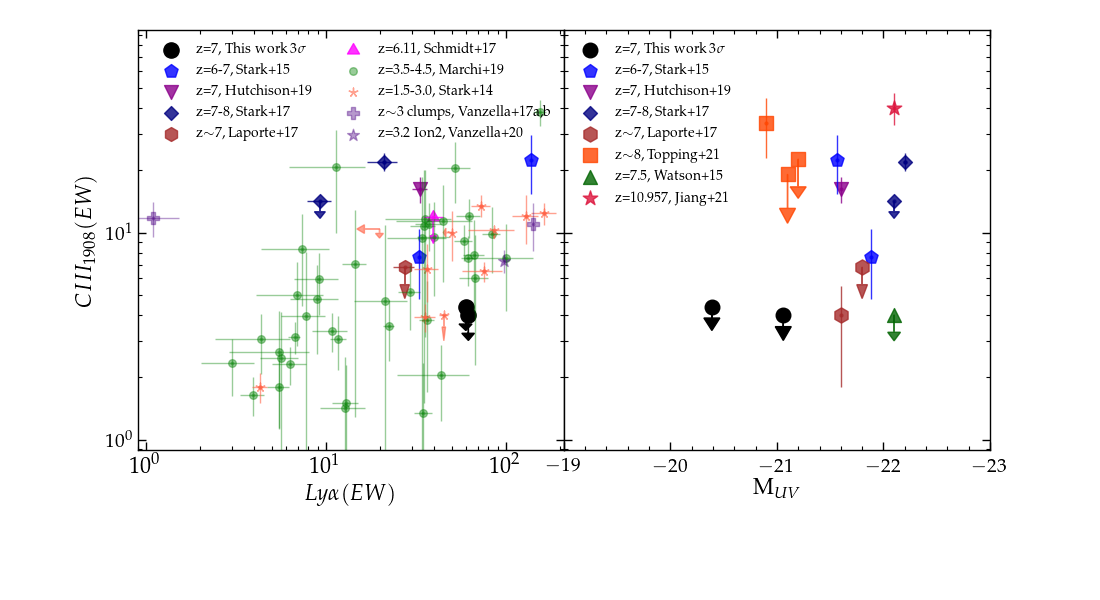}
\caption{The position of BDF521 and BDF2195 on the CIII](EW) versus Ly$\alpha$(EW) plane (left panel), and on the CIII](EW) vs $M_{UV}$ plane (right panel), compared to $z\gtrsim$6 objects from the literature: \citet{Stark2015a}, \citet{Watson2015}, \citet{Stark2017}, \citet{Laporte2017}, \citet{Schmidt2017}, \citet{Hutchison2019}, \citet{Jian2021}, \citet{Topping2021}. In the left panel we also show objects at $z\sim$1.5-3 from \citet{Stark2014}, the CIII] emitters at z=3.5-4.5 from the VANDELS survey \citep{Marchi2019}, the Lyman-continuum leaker Ion2 \citep[z=3.2,][]{Vanzella2020}, and the ultra-faint star-forming clumps at z$\sim$3 from \citet[][]{Vanzella2017a,Vanzella2017b}. All upper limits are at 3$\sigma$, see legend for symbols and colors.} \label{CIIIfig}
\end{figure*} 

\begin{figure*}
\centering
\includegraphics[width=\textwidth,height=\textheight,keepaspectratio]{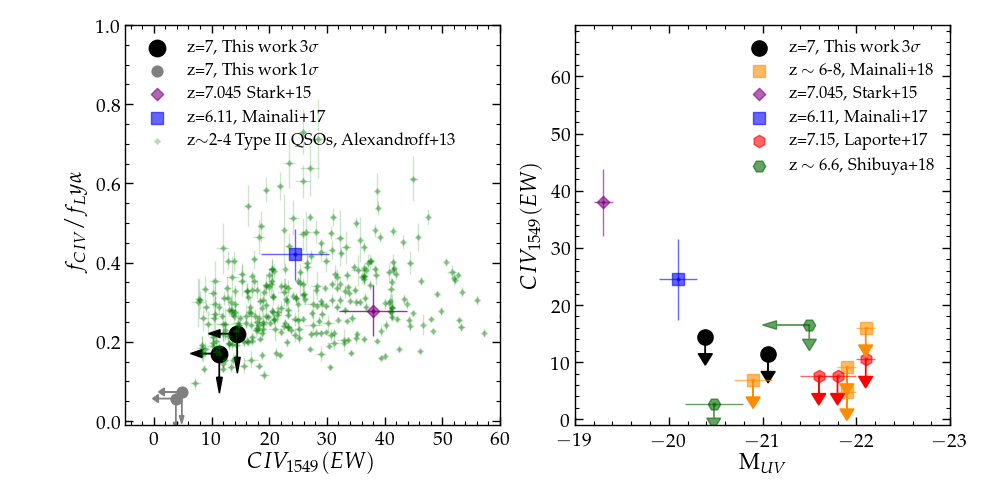}
\caption{\textbf{Left:} the position of BDF521 and BDF2195 on the $f_{CIV}/f_{Ly\alpha}$ versus Ly$\alpha$(EW) plane compared to $z\gtrsim$6 objects with detected CIV emission \citep[][]{Stark2015b,Mainali2017}, and to Type II QSOs at intermediate redshift from \citet{Alexandroff2013}. \textbf{Right:} the position of the BDF LAEs on the CIV(EW) versus $M_{UV}$ plane compared to $z\gtrsim$6 sources from the literature including upper limits from \citet{Laporte2017}, \citet{Mainali2018}, and \citet{Shibuya2018}. All upper limits are at 3$\sigma$, see legend for symbols and colors.} \label{CIVfig}
\end{figure*} 

\subsection{Limits on UV metal lines} \label{subsec:Lya}
We inspected both 2D and 1D spectra to assess the presence of metal UV emission lines. We detect no continuum emission and find no significant features within 500 km/s from the position expected on the basis of the Ly$\alpha$ redshift. In general, no evident line emission is elsewhere found at the objects' positions in the NIR arm spectra. Considering the Ly$\alpha$ velocity offset range expected for bright $z\sim$7 galaxies \citep[e.g.][]{Maiolino2015,Mason2018} we estimate upper limits for the flux of all lines from the average flux uncertainty in a moving window of 100 km/s rest-frame across the range between -500 km/s and 0 km/s from the Ly$\alpha$ redshift, while masking sky line residuals. We derive corresponding EW limits after estimating the continuum flux at each position on the basis of the observed photometry (C16). Namely, we follow the standard assumption of approximating the UV continuum as a power law $F_{\lambda}\propto\lambda^{\beta}$ whose slope $\beta$ can be measured from a regression across the observed bands \citep[e.g.][]{Castellano2012}. The measured slopes are $\beta=$-2.25$\pm$0.25 and $\beta=$-1.66$\pm$0.16 for BDF521 and BDF2195, respectively. We find rest-frame EW limits $\lesssim$2-7~\AA~at 3$\sigma$ (Table~\ref{linestable}). The limit on HeII emission for BDF521 is consistent with the estimate obtained by \citet[][]{Cai2015} from narrow-band HST photometry. The spectra at the expected positions of the relevant lines are shown in Figs.~\ref{linesBDF521} and \ref{linesBDF2195}.

\subsection{Comparison with other high-redshift sources} \label{subsec:comparison}

In the last few years, several groups actively searched for UV emission lines in the most distant galaxies, as valuable diagnostics of the underlying radiation field \citep[e.g.,][]{Mainali2018,Stark2017,Laporte2017,Endsley2021b} and in some case as redshift indicators, in the absence of Ly$\alpha$ \citep{Jiang2021}.
The most common line, besides Ly$\alpha$, is the CIII] doublet which has been detected in several of the most distant galaxies including the most distant spectroscopically confirmed galaxy GN-z11 at z=10.957 \citep{Oesch2016,Jiang2021}.
At $z\sim$3-4, a strong  correlation between the Ly$\alpha$ and CIII] strength has been found by several authors \citep[][]{Stark2014,Marchi2019}, although in other cases a large scatter between the quantities was also observed \citep[][]{LeFevre2019,Rigby2015,Llerena2021,Schmidt2021}. Confirmed \citep[][]{Vanzella2020} or suspected $z\sim$3 Lyman-continuum leakers \citep[][]{Vanzella2017a,Vanzella2017b} either fall on the expected relation, or rather show significant CIII] emission and suppressed Ly$\alpha$ emission. 

It is unclear whether the relation holds at z$\gtrsim$6 where only a few detections are available. The reported detections we show in the left panel of Fig.~\ref{CIIIfig} seem to agree with the lower redshift behavior,  while our two BDF emitters do not follow the expected trend. Indeed, the galaxies  have rather large Ly$\alpha$ EW and we would expect solid detections of the  CIII]  doublet at $>5\sigma$ from the average of lower redshift relations, albeit the BDF objects are consistent with the lower envelope of the measured distributions. As shown in the right panel of Fig.~\ref{CIIIfig}, all high-redshift objects with a secure CIII] detection are bright, $L\gtrsim L*$ LBGs \citep[i.e., $M_{UV}\lesssim$ -20.5, e.g.][]{Harikane2021}, in most cases significantly brighter than the BDF emitters.

\citet[][]{Llerena2021} found that strong  CIII] emitters at $z\sim$3 have strong radiation fields, low stellar metallicity but their gas is already partially enriched with C/O abundances in the range 35\%-150\% solar. In this respect, the lack of strong CIII] emission can be explained either by an extremely low C abundance in the gas phase, and/or by the objects being in a less active star-formation phase \citep[][]{Berg2019}. In the case of the BDF galaxies a low metallicity is likely the main factor affecting the low CIII] emission given the high Ly$\alpha$ EW suggesting active star-formation. A low metallicity is also consistent with the lack of CII158$\mu m$ detection in BDF521 \citep[][]{Maiolino2015}. A strong starburst phase that photodissociates the molecular clouds could also lower CII158$\mu m$ emission \citep[][]{Pallottini2019}, but UV CIII] emission should be enhanced in such a situation if C is abundant in the ionized gas.
\\
In Fig.~\ref{CIVfig} we compare the limits we have obtained for the CIV emission on the two BDF galaxies to previous studies. In the EoR there are only two  galaxies where  a convincing CIV emission has been detected, the gravitationally lensed low-mass galaxies RXC J2248.7-4431-ID3 \citep[z=6.11,][]{Mainali2017}, and A1703-zd6 \citep[z=7.045,][]{Stark2015b}. Several attempts to secure this line in small samples of galaxies at z=6-10 have failed to report detections  \citep[][]{Laporte2017,Mainali2018,Shibuya2018}, with typical EW limits around 3-10 \AA~comparable to the BDF galaxies. Although an AGN contribution to the Ly$\alpha$ flux cannot be completely ruled out, the CIV equivalent widths and $f_{CIV}/f_{Ly\alpha}$ ratios for the two BDF galaxies are low compared to most known Type II AGN at intermediate redshift \citep[][]{Alexandroff2013}, and much lower than those for the two $z>$6 LBGs with CIV detections (left panel of Fig.~\ref{CIVfig}). These results are consistent with the analysis of the Ly$\alpha$/NV ratio from FORS2 spectroscopy (C18). The two BDF galaxies have Ly$\alpha$/NV$\gtrsim$8-10 which is higher than the values measured in suspected AGN at z$\gtrsim$7 \citep[e.g.,][]{Tilvi2016,Laporte2017,Mainali2018}, albeit still consistent with values observed in AGN at lower redshifts \citep{Humphrey2008,Hainline2011}.

\section{Limits on the contribution from AGN emission} \label{sec:AGN}
\begin{figure}
\centering
\includegraphics[trim={0.1cm 0.1cm 0.1cm 0.1cm},clip,width=\linewidth,keepaspectratio]{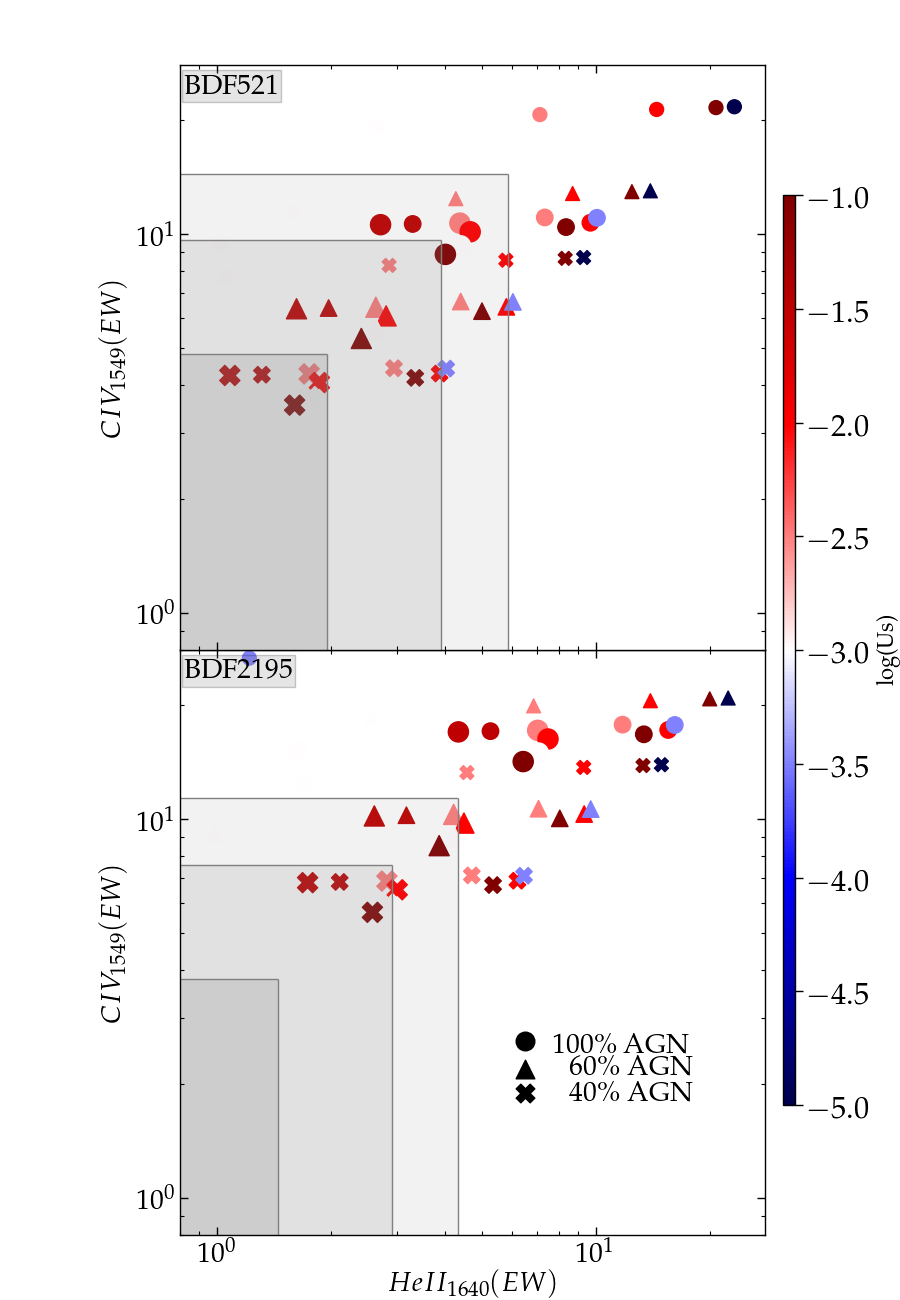}
\caption{The position in the CIV(EW) versus HeII(EW) plane of models with 40\% (crosses), 60\% (triangles) and 100\% (circles) contribution to the observed UV emission of the two BDF galaxies from narrow-line AGN \citep[from][]{Feltre2016} with different metallicities and ionization parameter. Symbols at increasing size are for models with Z=0.025, 0.1, 0.4 $Z_{\odot}$, whose color indicate the relevant ionization parameter log($U_{s}$). The shaded regions highlight the parameter space allowed at 1$\sigma$, 2$\sigma$ and 3$\sigma$ by XSHOOTER observations of BDF521 (top) and BDF2195 (bottom).} \label{AGNfig}
\end{figure} 

The analysis of the limits on CIV emission presented in Fig.~\ref{CIVfig}, and the existing constraints on the Ly$\alpha$/NV ratio (C18), already suggest that it is unlikely that the BDF emitters are dominated by AGN emission.  

We perform here a more quantitative analysis by comparing the limits on CIV and HeII emission against narrow line AGN models by \citet{Feltre2016}.
We consider models with -5.0 $\leq$ log($U_{s}$) $\leq$ -1.0 and interstellar gas metallicity $Z=0.025, 0.1, 0.4 ~Z_{\odot}$, dust-to-metal mass ratio $\xi_d=0.3$, and hydrogen gas density $n_H$=100~cm$^{-3}$. We restrict our analysis to models generated with incident spectra having UV spectral index $\alpha=-1.7$, but we verified that results do not significantly change when adopting spectra with index $\alpha=-1.4$. We first re-normalize the incident spectra to a chosen fraction of the observed flux of each object in the F125W filter (C16), i.e., the observable non-ionizing UV flux at $\sim 1500$\AA, and then compute the EW of the CIV and HeII lines on the basis of the predicted line flux for each model and of the observed flux at the relevant wavelength. We note that using the models in this way allows us to constrain scenarios in which the AGN is contributing to the UV continuum emission, which is the most important test to constrain the ionizing output of our targets.  While the models do not include the broad line region emission, this is not a concern in our case given the narrow FWHM of the Ly$\alpha$ lines observed in the BDF galaxies. We show in Fig.~\ref{AGNfig} the position on the CIV(EW) versus HeII(EW) plane of AGN models with a 40\%, 60\% and 100\% contribution to the observed UV emission highlighting the regions allowed at 1$\sigma$, 2$\sigma$ and 3$\sigma$ by the XSHOOTER observations.
Models with  pure AGN (100\%) are always excluded at $\gtrsim2-3\sigma$ regardless of log($U_{s}$) and metallicity. The more stringent constraints on BDF2195 allow us also to exclude a 60\% AGN contribution at $\sim$3$\sigma$.
Models with 40\% contribution are more compatible with the observations, particularly for BDF521, for high log($U_{s}$) and 
intermediate/high metallicities.

The narrow line models of \cite{Feltre2016} can be further used to investigate the case of an obscured AGN providing no contribution to the UV continuum emission. In such a case, we must consider the fractional contribution of the AGN to the observed Ly$\alpha$ emission.  Since the radiative transfer effects of Ly$\alpha$ are severe, we therefore investigate what fraction of Ly$\alpha$ can escape for given fractional contribution of AGN without invalidating the measured line limits.  For a similar parameter space of the models investigated above, we find that we can only rule out 100\% narrow-line region contribution to Ly$\alpha$ if only 20\% (15\%) escapes for BDF2195 (BDF521).  For lower escape fraction of Ly$\alpha$, the line contribution can have 100\% contribution from a Type-II AGN without the other rest-frame UV lines being detected.

\section{The ionizing emission budget of the BDF emitters}\label{sec:ion}
From the above analysis it appears that the BDF pair lacks high-ionization features found in some other sources at a similar redshift, and instead resembles standard star-forming galaxies at lower redshifts with low or absent contribution from AGN. 
We thus constrain the physical parameters of the two galaxies, including the ionizing budget, by means of a spectro-photometric fit performed with the \textsc{BEAGLE} v0.24.5 tool \citep{Chevallard2016} using the most recent version of the \citet[][]{Bruzual2003} stellar population synthesis models \citep[see][for details]{VidalGarcia2017}. Nebular emission is modeled self-consistently as described in \citet[][]{Gutkin2016} by processing stellar emission with the photoionization code \textsc{CLOUDY} \citep[c13.03,][]{Ferland2013}. The fit is performed by fitting the upper limits on the integrated lines plus continuum fluxes measured as described in Sect.~\ref{sec:lines}, together with the broad-band photometric measurements redward of the Lyman break in the F105W, F125W, F160W and HAWK-I Ks bands \citep[C16, see also][]{Castellano2010b,Cai2015}. 
Given that the Ly$\alpha$ is very difficult to model in this framework, we subtract its flux from the affected bands and do not include it in the nebular emission modeled by \textsc{BEAGLE}. We fix the redshift at the value determined from Ly$\alpha$. Although there is usually a shift between Ly$\alpha$ and systemic redshift, this in practice  has no effect on the constraints derived here. The \textsc{BEAGLE} SED-fitting runs are performed with a configuration similar to the one discussed by \citet{Stark2017}. The templates are based on a \citet{Chabrier2003} initial mass function and have metallicity in the range $-2.2 \leq log(Z/Z_{\odot}) \leq 0.25$. The star formation histories (SFH) are parametrized as an exponentially delayed function (SFR(t) $\propto$ t$\cdot$exp(-t/$\tau$)), which is the most flexible parametric SFH allowed by the code, plus an ongoing constant burst of 10 Myr duration. We adopt uniform priors on the SFH exponential timescale (7.0 $\leq$ $\tau$/log(yr) $\leq$ 10.5), stellar mass (7.0 $\leq$ log($M/M_{\odot}$) $\leq$ 12), and maximum stellar age (7.0 $\leq$ log($Age/yr$) $\leq$  ~age of the universe). Attenuation by dust is treated following the \citet{Charlot2000} model combined with the \citet{Chevallard2013} prescriptions for geometry and inclination effects, assuming an effective V-band optical depth in the range -3.0 $\leq$ log($\tau_{V}$) $\leq$ 0.7 with a fixed fraction $\mu=0.4$ arising from dust in the diffuse ISM. Interstellar metallicity $Z_{ISM}$ is assumed to be identical to the stellar one, the dust-to-metal mass ratio and ionization parameter are left free in the ranges $0.1 \leq \xi_d \leq 0.5$ and -4.0 $\leq$ log($U_{s}$) $\leq$ -1.0, respectively.

We exploit the observed Ly$\alpha$ flux to impose a prior on the ongoing SFR. Following \citet{Kennicutt1998}, the total line fluxes correspond to SFR(Ly$\alpha$)=6.7$\pm$0.5$~M_{\odot}$/yr and SFR(Ly$\alpha$)=11.4$\pm$0.6$~M_{\odot}$/yr for BDF521 and BDF2195, respectively. As we expect the SFR inferred from the detected Ly$\alpha$ to be a lower limit to the intrinsic SFR of BDF521 and BDF2195 depending on their Ly$\alpha$ escape fraction and attenuation by the IGM, we use the corresponding 5$\sigma$ lower limits on SFR(Ly$\alpha$), i.e. $\simeq4~M_{\odot}$/yr and $\simeq 8~M_{\odot}$/yr, as conservative lower bounds for the uniform prior ranges on their current SFRs.

The best-fit parameters and relevant 68\% confidence level uncertainties are shown in Table.~\ref{beagletable}. The objects are found to be relatively young (mass-weighted age $\sim$20-30 Myrs) and metal-poor ($\lesssim 0.3 Z_{\odot}$) galaxies with stellar masses of a few $10^9M_{\odot}$. Both objects are actively star-forming (SFR$\sim$15$M_{\odot}/yr$), but had experienced a higher star-formation activity in the past. We find a small degeneracy among the mass, extinction and age parameters, with older models being more massive and less extincted by dust. The results do not change significantly when adopting different parametrizations of the SFH, nor with updated treatment of dust within HII regions implemented in BEAGLE v0.27.1 described in \citet{Curtis-Lake2021}.

The most important parameter to explore thanks to the combined spectroscopic and photometric information is the production rate of hydrogen ionizing photons per intrinsic (i.e. unattenuated) UV luminosity, $\xi_{ion}^*$. We find best-fit values for log($\xi_{ion}^*$/Hz erg$^{-1}$) of 25.26 and 25.02 for BDF521 and BDF2195, respectively. These values are consistent with the typical ionizing production efficiency of $z\gtrsim$4 populations  \citep[][]{Bouwens2015b,Bouwens2016a,Lam2019}, and significantly lower than the log($\xi_{ion}^*$/Hz erg$^{-1}$)$\simeq$25.6-25.7 found in $z>$7 galaxies with CIII] detection and/or photometric evidence of strong optical emission lines \citep[][]{Stark2015b,Stark2017}. The ionizing emission from the BDF emitters is lower, albeit consistent at $\sim1\sigma$, with the average value in $z\sim$5 LAEs \citep[][]{Harikane2018}, and with the log($\xi_{ion}^*$/Hz erg$^{-1}$)=25.43 estimated for the BDF galaxies by \citet{Espinosa2021} on the basis of the Ly$\alpha$ luminosity following \citet{Sobral2019}. 
Most importantly, these results show that the two emitters are rather "standard" high-redshift star-forming galaxies, with no significant excess production of ionizing photons contributing to the local reionization history.

\begin{table}
\caption{Properties of the BDF pair$^a$}
\centering
\begin{tabular}{ c | l | l |}
%\hline
\cline{2-3}
   & BDF521 & BDF2195\\
%\cline{2-3}
\hline 
\multicolumn{1}{|c|}{Redshift} & 7.0121 & 7.0124 \\
\hline 
\multicolumn{1}{|c|}{$M_{UV}$} & -20.4& -21.1\\
\hline 
\multicolumn{1}{|c|}{UV slope} & -2.25$\pm$0.25  & -1.66$\pm$0.16\\
\hline 
\multicolumn{1}{|c|}{$M_{star}~(10^9 M_{\odot})$} &
$1.04^{+0.48}_{-0.40}$ & $2.67^{+1.11}_{-1.19}$\\
\hline 
\multicolumn{1}{|c|}{SFR ~($M_{\odot}$/yr)}&$14.9^{+4.8}_{-5.4}$  &$16.0^{+9.5}_{-9.8}$  \\
\hline 
\multicolumn{1}{|c|}{Age$^b$~(Myr)} & $23^{+5}_{-12}$  &$34^{+20}_{-20}$\\
\hline 
\multicolumn{1}{|c|}{Stellar $Z/Z_{\odot}^b$} & $0.14^{+0.15}_{-0.13}$ &$0.12^{+0.11}_{-0.11}$\\
\hline 
\multicolumn{1}{|c|}{$A_{1500}$ (mag)}&$1.4^{+0.40}_{-0.40}$ & $1.4^{+0.38}_{-0.40}$\\
\hline 
\multicolumn{1}{|c|}{$\xi_{ion}^*$~ (log(erg/Hz))}&$25.26^{+0.1}_{-0.1}$  &$25.02^{+0.22}_{-0.22}$\\
\hline
\end{tabular} \label{beagletable}
\small \\ $^a$ Physical properties (mean values and 68\% c.l. uncertainties) are from the BEAGLE spectro-photometric fit. 
\\$^b$ Mass-weighted values.
\end{table}

\begin{figure*}
\centering
\includegraphics[trim={0.5cm 0.3cm 0.5cm 0.3cm},width=\linewidth,keepaspectratio]{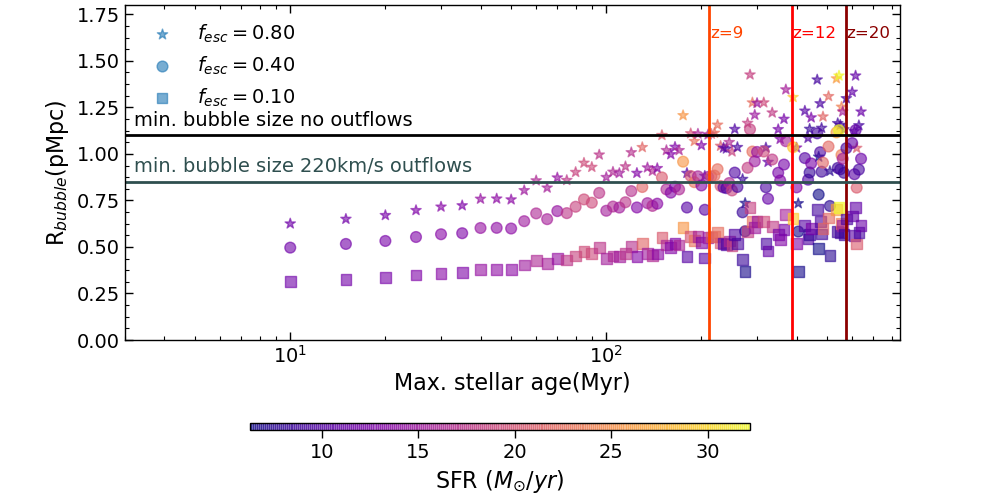}
\caption{The size ($R_{bubble}$) of the ionized bubble created by the combined emission of the BDF pair, as a function of the maximum stellar age of the stellar population for BEAGLE SED models within 95\% c.l. from the best-fit. The cases for escape fraction $f_{esc}$=0.1,0.4, 0.8 are shown as squares, circles and stars, respectively, with color indicating the relevant, total, SFR of the system. The horizontal black line marks the minimum HII size $R_{min}$=1.1 Mpc enabling Ly$\alpha$ to escape. The dark grey line indicates the minimum HII size  $R_{min}$=0.85 Mpc in the case Ly$\alpha$ is affected by a 220km/s outflow. Vertical lines, from left to right, indicate formation redshifts z=9, 12 and 20.} \label{bubblefig}
\end{figure*} 

With the improved constraints coming from the new spectro-photometric fit described in the previous section, we can quantitatively explore whether the BDF pair is capable of re-ionizing its surroundings or not. The SFR and the age of a galaxy can be used to measure the size ($R_{bubble}$) of the resulting ionized region following, e.g., \citet{Shapiro-giroux1987} and \citet{Madau1999} by assuming an escape fraction of ionizing photons $f_{esc}$, the hydrogen clumping factor C, and an average neutral hydrogen fraction $\chi_{HI}$ at the onset of star-formation. As discussed in V11 and C18, the size $R_{bubble}$ must then be compared to the minimum size $R_{min}$=1.1 Mpc enabling Ly$\alpha$ to be redshifted enough to reach the observers \citep{Wyithe2005}. The aforementioned $R_{min}$ is found under the assumption that the Ly$\alpha$ escapes from the galaxies at the systemic redshift. The size threshold decreases in presence of strong outflows, for example  a 220km/s shift, which is the median value found by \citet{Mason2018} \citep[see also][]{Endsley2022b} for galaxies in massive halos, results in $R_{min}\sim 0.85$ Mpc.

In practice, we compute the average SFR of the pair from all models within 95\% c.l. from the best-fit as a function of time in bins of 5 Myr. We then convert this into $R_{bubble}$ by assuming C=2 and $\chi_{HI}=0.5$ and let the escape fraction $f_{esc}$ vary from 10\% to 80\%. The assumption of a 50\% neutral fraction is conservative as a higher $\chi_{HI}$ would further decrease the size of the reionized region created by the sources. The results are shown in Fig.~\ref{bubblefig} as a function of the relevant maximum stellar age of the models (i.e., the onset of the star-formation episode): we find that in the absence of significant outflows only models with an extremely high $f_{esc}$=80\% and age larger than 200 Myr (formation redshift $z>$9) can create a large enough bubble. 

When assuming a  Ly$\alpha$ shift imprinted by 220km/s outflows, the two objects could grow a large enough bubble with an escape fraction $f_{esc}$=40\% on the same time frame. However, even in the presence of outflows, there are no spectro-photometric fitting solutions allowing the BDF pair to reionize the surrounding region with an escape fraction of $f_{esc}$=10\%. This is particularly relevant when considering that the physical properties of the BDF galaxies, i.e. moderate $\xi_{ion}^*$, and low EW of the UV metal lines, are typical for galaxies with low $f_{esc}$ \citep[][]{Naidu2022}. 
Note that these results are considerably  more stringent than those discussed in C18 based on the SED-fitting of the available photometry only. The C18 results were consistent with a combination of high SFRs and old ages which in turn allowed the formation of a large enough reionized bubble. Our current spectro-photometric analysis suggests that the two emitters are not solely responsible of the growth of the ionized bubble. In turn, this finding is consistent with the analysis of the BDF region presented by \citet{Espinosa2021} indicating that a dominant contribution is provided by the clustered, faint galaxies.

\section{Summary and future prospects}\label{sec:summary}

We have presented the analysis of deep VLT-XSHOOTER observations of BDF521 and BDF2195, a pair of $L \sim L^*$, bright Ly$\alpha$ emitting galaxies at $\sim$90 kpc separation shining within the BDF bubble, a $z\sim$7 overdensity of both LBGs and LAEs showing all expected properties of a "reionized bubble" embedded in a half-neutral universe \citep[V11, C16, C18,][]{Espinosa2021}. Our main findings can be summarised as follows:

\begin{itemize}
    \item The Ly$\alpha$ lines are detected at high significance in the VIS arm spectra at z=7.0121 and z-7.0124 for BDF521 and BDF2195, respectively. The Ly$\alpha$ properties (EW$\simeq$60\AA~and FWHM$\simeq$250-260km/s) are consistent with previous analysis based on lower resolution FORS2 spectra. 
    \item No significant emission is detected in the NIR arm spectra at the expected position of the CIV$\lambda 1548$ doublet, HeII$\lambda 1640$, OIII]$\lambda 1660$ doublet, and CIII]$\lambda 1909$ doublet. We can place stringent 3$\sigma$ upper limits on the EW of each line or doublet component in the range $\sim$2-7\AA~(Table \ref{linestable}).
    \item A comparison with other high-redshift sources from the literature shows that our two BDF emitters have lower CIII] emission than expected on the basis of the average correlation between the Ly$\alpha$ and CIII] EWs measured at lower redshifts, albeit they are consistent with being at the lower envelope of the measured distributions (Fig.~\ref{CIIIfig}). 
    \item Although an AGN contribution cannot be completely ruled out, the constraints on the CIV(EW) and the flux ratio $f_{CIV}/f_{Ly\alpha}$ (Fig.~\ref{CIVfig}) show that the objects are only marginally consistent with them being AGN when compared to known Type II QSOs at lower redshifts. Consistently, when comparing limits on HeII(EW) and CIV(EW) to narrow line AGN models by \citet{Feltre2016} we can exclude pure AGN (100\%) templates at $\sim2-3\sigma$. Instead, models with a $\lesssim$40\% AGN contribution could be compatible with the observations (Fig.~\ref{AGNfig}).
    \item A spectro-photometric fit on the available data indicate that the two objects are relatively young ($\sim$20-30 Myrs) and metal-poor ($\lesssim 0.3 Z_{\odot}$) with stellar masses of a few $10^9M_{\odot}$ (Table~\ref{beagletable}). Most importantly, we find a production rate of hydrogen ionizing photons per intrinsic UV luminosity of log($\xi_{ion}^*$/Hz erg$^{-1}$)=25.02-25.26 (under the assumption of a purely stellar origin), which is consistent with the value typically found in high-redshift populations, but significantly lower than the extreme values ($\simeq$25.6-25.7) measured in $z>$7 galaxies with CIII] detection \citep[][]{Stark2015b}  or with evidence of strong optical emission lines from their mid-IR photometry \citep[][]{Stark2017}.
    \item The range of physical parameters allowed by the spectro-photometric fit does not allow the BDF pair to reionize their surroundings by themselves with a low or even a moderate escape fraction of ionizing photons, as they would require extremely high $f_{esc}$=40\% and age larger than 200 Myr (formation redshift $z>$9) (Fig.~\ref{bubblefig}).
\end{itemize}

The results summarised above indicate that the two emitters are typical high-redshift sources with no peculiar ionizing capabilities, either due to their stellar populations or to a significant contribution from AGN emission. The low metal line emission and the moderate $\xi_{ion}^*$ are likely explained by a combination of low metallicity and the galaxies being caught in a quiet phase of star-formation activity. In addition, the lack of a significant AGN contribution is not surprising in the light of theoretical predictions showing a dominant contribution from star-formation to the UV emission of bright high-redshift galaxies \citep[][]{Piana2022}. Most importantly for our purposes, it is unlikely that they are solely responsible of the growth of the ionized bubble. Instead, a dominant contribution to the local reionization history can be provided by the overdensity of faint galaxies. 

These findings allow us to draw a plan for future observations capable of fully constraining the properties of this remarkable region. As of today, the most important missing ingredient to confirm that the  faint galaxies in the overdensity reionized the BDF region is their spectroscopic confirmation. Given the absence of strong Ly$\alpha$ this is beyond the capabilities of ground-based telescopes, while attempting detection of submillimeter lines with ALMA in redshift scan mode would be exceedingly time consuming. However, spectroscopic confirmation is well within reach of the James Webb Space Telescope. Indeed, JWST-NIRSpec in $\sim$1.5 hours of integration time could detect [OIII]$\lambda 4959,5007$ from all currently known candidates down to $M_{UV}\simeq -19$, while also fully constraining ionization state and metallicity of the LAEs through the detection of [OII]$\lambda 3727$ and H$\beta$. 
In addition, JWST can extend the photometric mapping of the BDF in terms of area, depth and wavelength coverage to confirm the presence of an overdensity of LBGs down to fainter magnitudes. For example, $\sim$1-2 hours of NIRCam integration per filter can extended the constraints on the local UV LF down to $M_{UV}\sim -18.0$, while covering the optical rest-frame wavelength range which is crucial to reject low redshift interlopers and constrain the spectral energy distributions. 

Finally, a significant contribution to our understanding of the formation of reionized regions will be provided by objects beyond the limit that can be reached on blank fields such as the BDF, e.g. objects analogous to the $z\sim$6 clustered ultra-faint dwarfs observed by \citet{Vanzella2019} \citep[see also][]{Vanzella2021}. In this respect, JWST surveys on lensed fields \citep[][]{Treu2017prop,Willott2017prop,Vanzella2021prop}  will provide crucial information, such as an accurate characterization of the faintest end of the UV LF and of its possible turn-over \citep[e.g.,][]{Castellano2016c,Yue2018}, and of the physical properties of high-redshift ultra-faint dwarfs \citep[e.g.,][]{Vanzella2017a,Vanzella2017b}.

With the sample of candidate reionized regions steadily growing, the advent of JWST will enable a systematic investigation of the connection between the physical properties of high-redshift galaxies and the ionization state of the surrounding IGM, eventually providing an answer on which are the sources of reionization.

\begin{acknowledgements}
Based on observations collected at the European Southern Observatory
for Astronomical research in the Southern Hemisphere under
ESO programme 0103.A-0710. We thank A. Feltre for the useful discussions and support on the use of the narrow-line AGN models. ECL acknowledges support of an STFC Webb Fellowship (ST/W001438/1). EV acknowledges funding from the INAF  for ``interventi aggiuntivi a sostegno della ricerca di main-stream'' and
PRIM-MIUR 2017WSCC32 ``Zooming into dark matter and proto-galaxies with massive lensing clusters''. RA acknowledges support from ANID FONDECYT Regular Grant 1202007. PD and AH acknowledge support from the European Research Council's starting grant ERC StG-717001 (\quotes{DELPHI}). PD acknowledges support from the NWO grant 016.VIDI.189.162 (\quotes{ODIN}) and the European Commission's and University of Groningen's CO-FUND Rosalind Franklin program. 
A. Ferrara and S. Carniani acknowledge support from the ERC Advanced Grant INTERSTELLAR H2020/740120. Partial support from the Carl Friedrich von Siemens-Forschungspreis der Alexander von Humboldt-Stiftung Research Award is kindly acknowledged. This research made use of the \textsc{Matplotlib} package\footnote{https://matplotlib.org/} \citep{Hunter2007}, of \textsc{Astropy}\footnote{http://www.astropy.org}, a community-developed core Python package for Astronomy \citep{astropy:2013, astropy:2018}, and of the \textsc{Specutils} package\footnote{https://specutils.readthedocs.io/en/stable/}.
\end{acknowledgements}

\bibliographystyle{aa}

\appendix
\newpage
\section{Observed spectra}\label{sec:appendix}
\begin{figure*}[!t]
\centering
\includegraphics[trim={1.8cm 0.5cm 1.5cm 0.5cm},width=6.0cm]{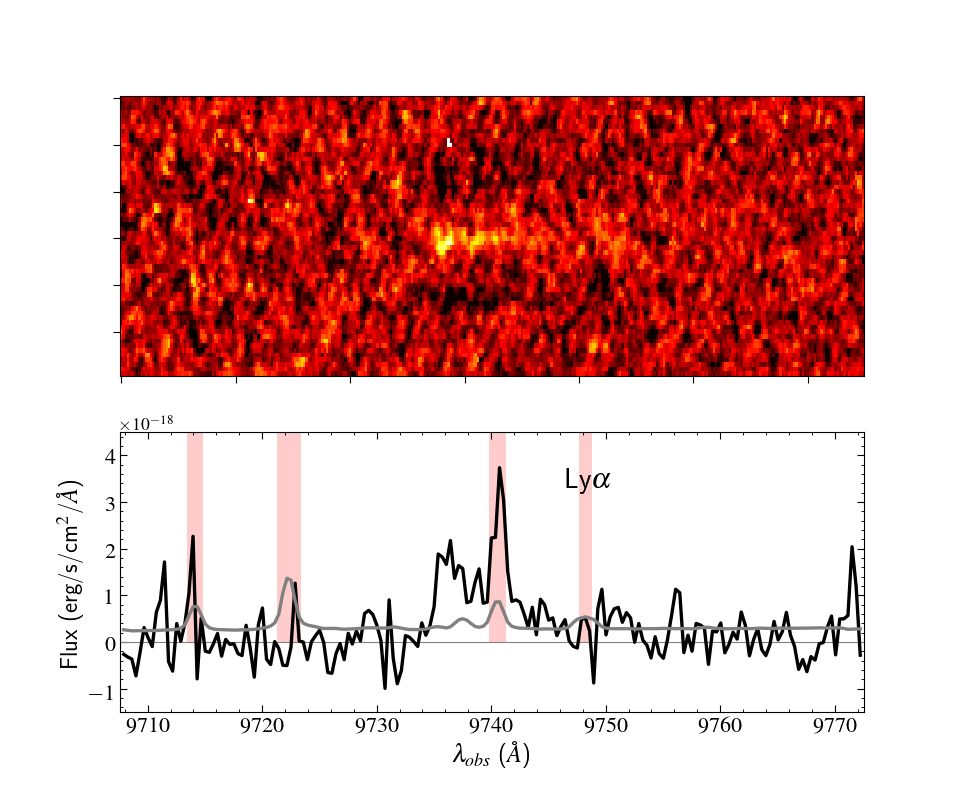}
\includegraphics[trim={1.8cm 0.5cm 1.5cm 0.5cm},width=6.0cm]{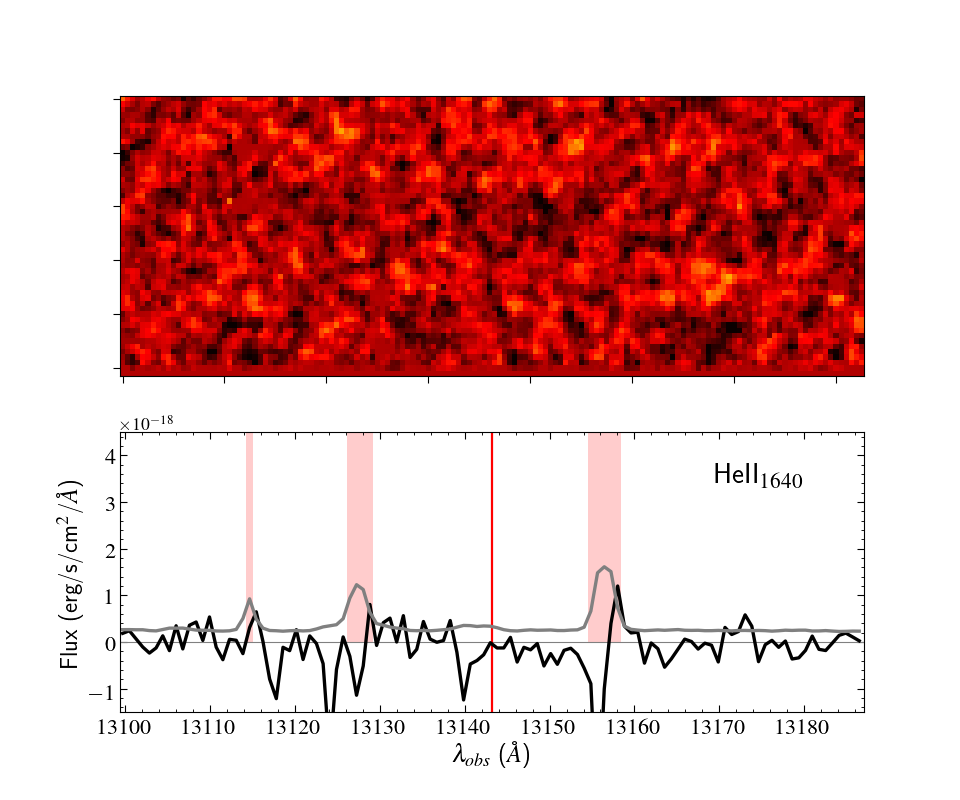}
\includegraphics[trim={1.8cm 0.5cm 1.5cm 0.5cm},width=6.0cm]{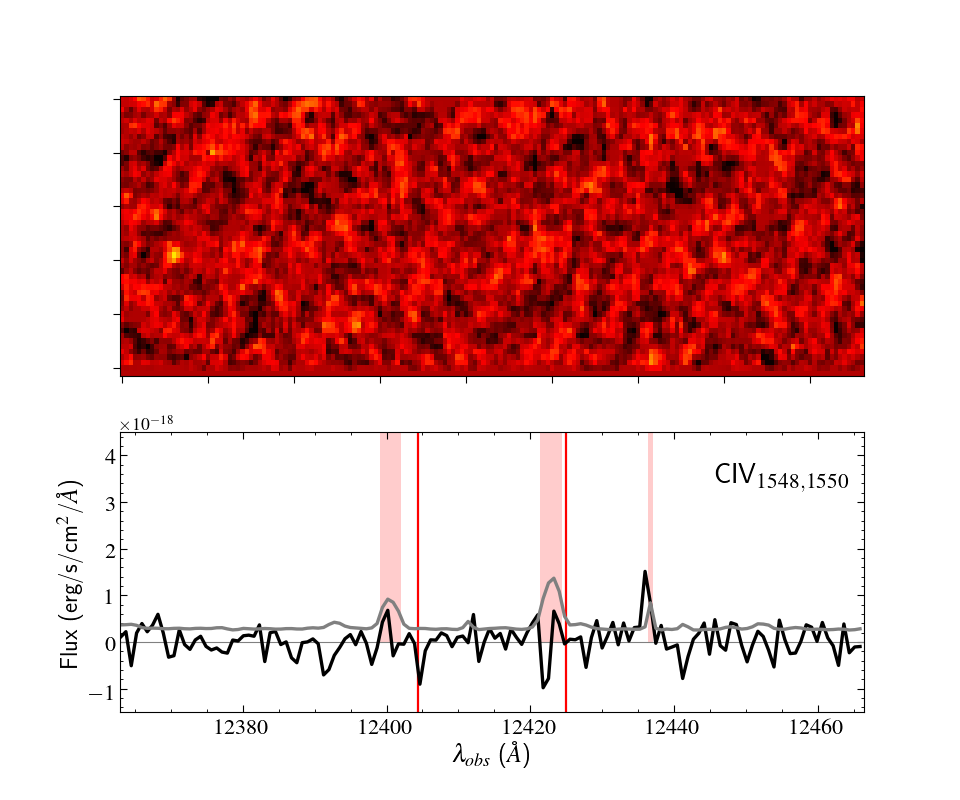}
\includegraphics[trim={1.8cm 0.5cm 1.5cm 0.5cm},width=6.0cm]{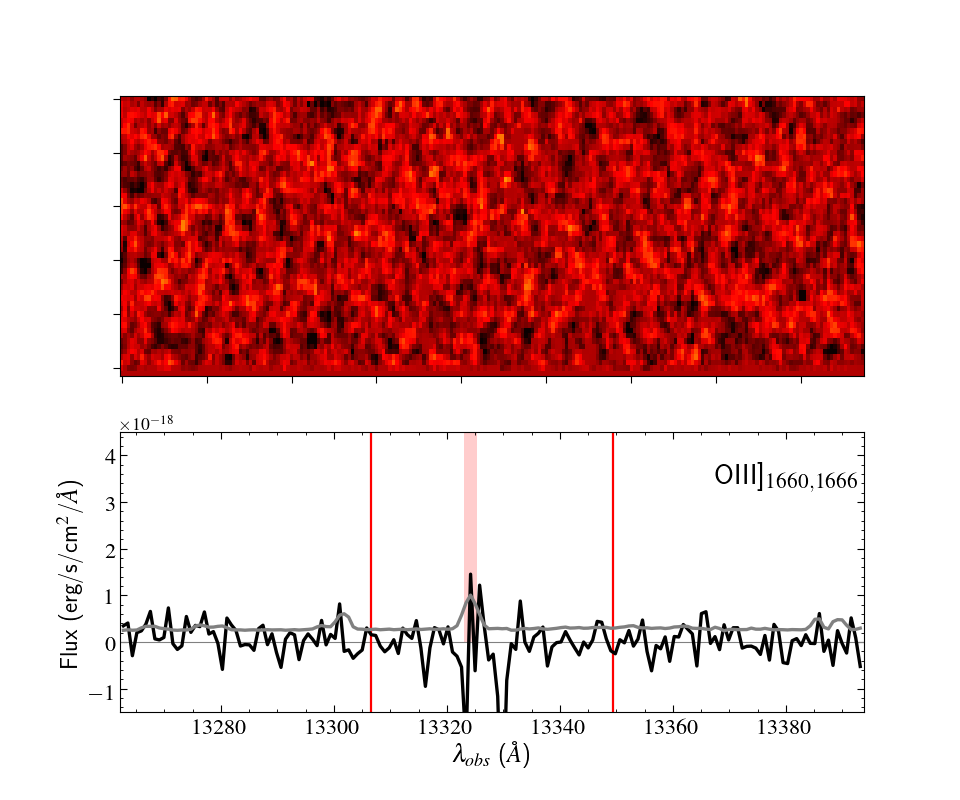}
\includegraphics[trim={1.8cm 0.5cm 1.5cm 0.5cm},width=6.0cm]{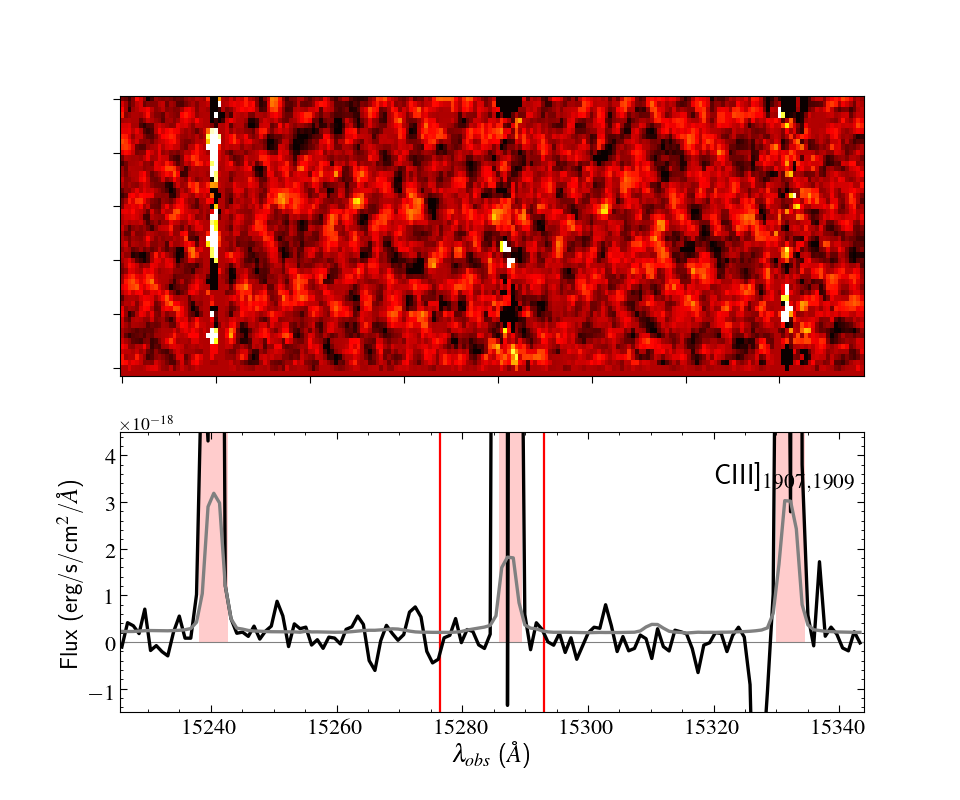}
\caption{Observed spectral regions at the position of BDF521 UV lines observed by X-SHOOTER. The regions cover the range from -1000 to +1000 km/s around the Ly$\alpha$ redshift, the red vertical lines mark the expected positions of UV lines or doublet components assuming no velocity shift with respect to the Ly$\alpha$. Top panels show the 2D S/N spectrum, while bottom panels show the 1D spectrum and rms as black and grey lines, respectively.  Shaded red regions are masked due to contamination from sky lines.}\label{linesBDF521}
\end{figure*}
\begin{figure*}
\centering
\includegraphics[trim={1.8cm 0.5cm 1.5cm 0.5cm},width=6.0cm]{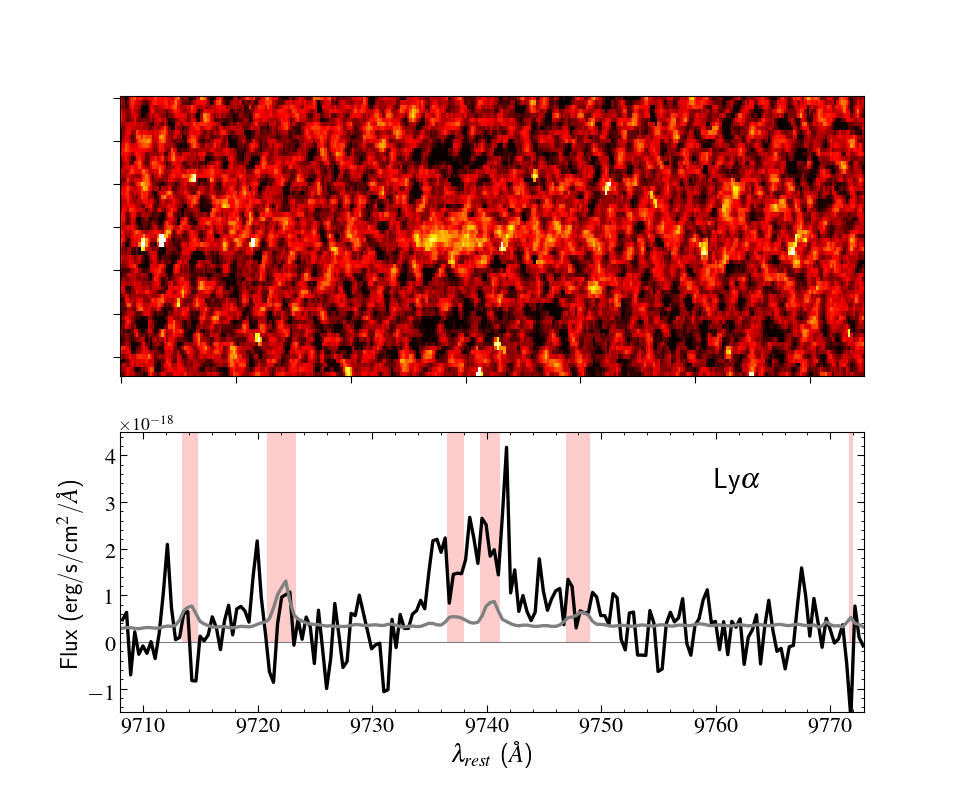}
\includegraphics[trim={1.8cm 0.5cm 1.5cm 0.5cm},width=6.0cm]{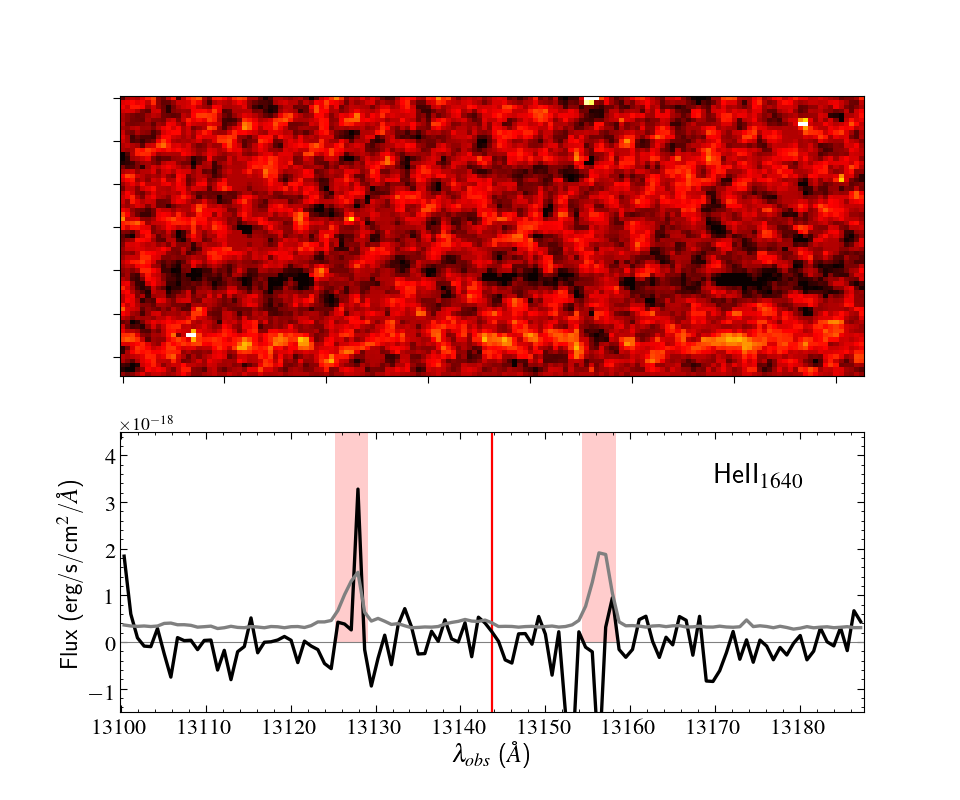}
\includegraphics[trim={1.8cm 0.5cm 1.5cm 0.5cm},width=6.0cm]{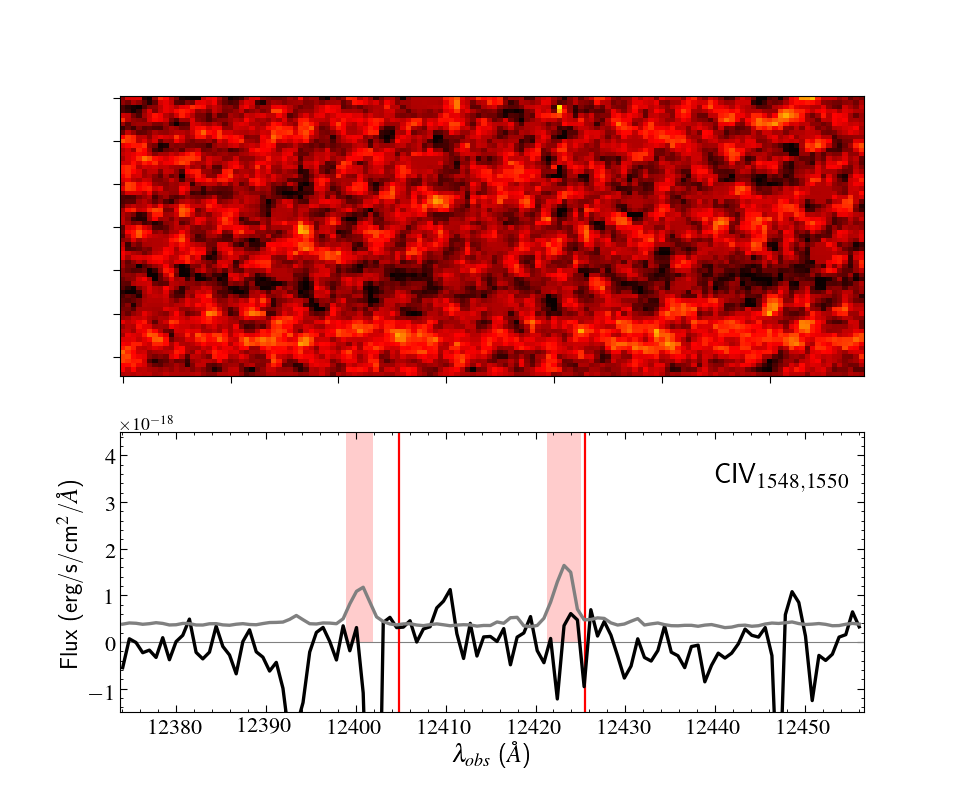}
\includegraphics[trim={1.8cm 0.5cm 1.5cm 0.5cm},width=6.0cm]{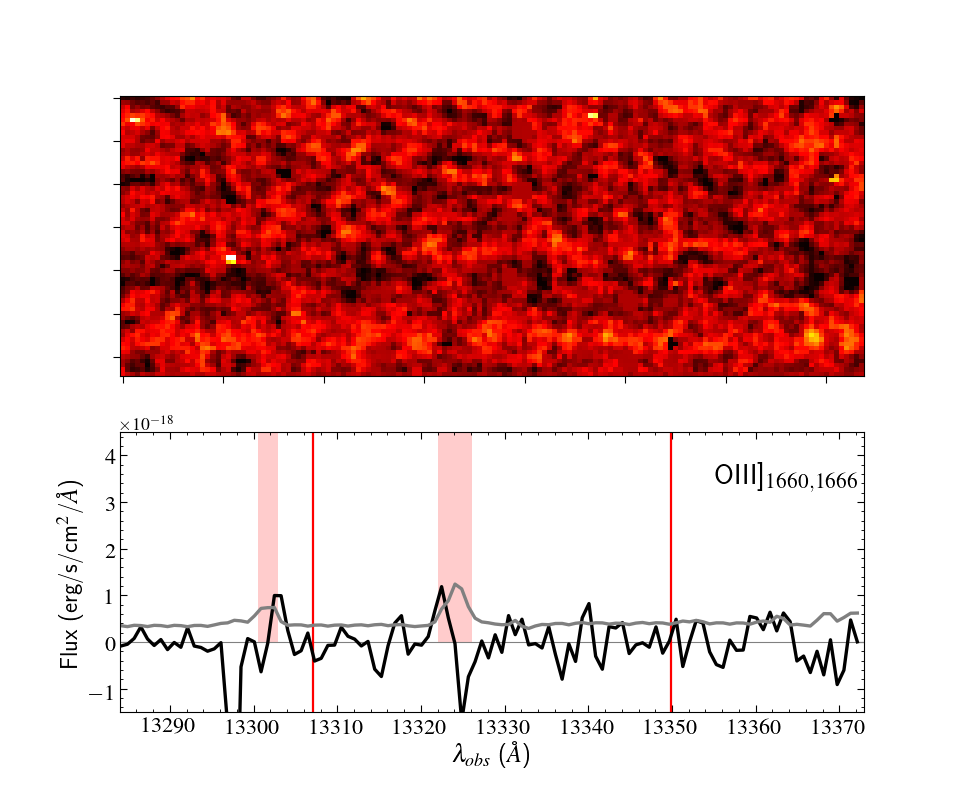}
\includegraphics[trim={1.8cm 0.5cm 1.5cm 0.5cm},width=6.0cm]{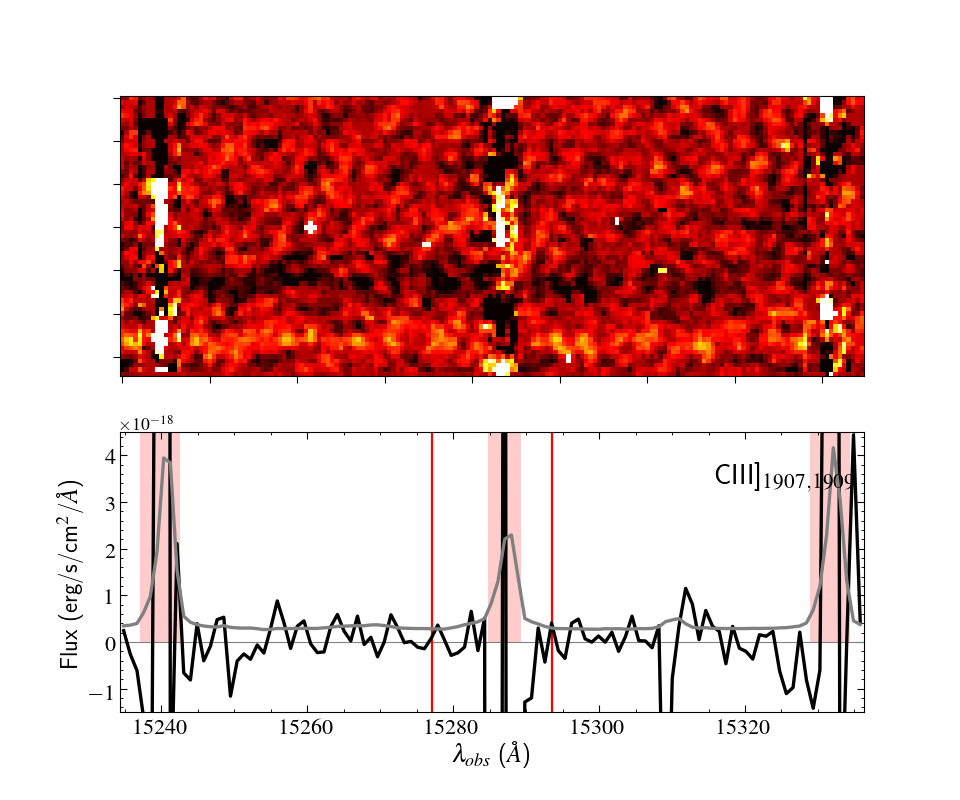}
\caption{Same as ~\ref{linesBDF521} for BDF2195.}\label{linesBDF2195}
\end{figure*} 
\end{document}